\documentclass{article}

\usepackage{PRIMEarxiv}

\usepackage[utf8]{inputenc} 
\usepackage[T1]{fontenc}    
\usepackage{hyperref}       
\usepackage{url}            
\usepackage{booktabs}       
\usepackage{amsfonts}       
\usepackage{nicefrac}       
\usepackage{microtype}      
\usepackage{lipsum}
\usepackage{fancyhdr}       
\usepackage{graphicx}       
\graphicspath{{media/}}     

\usepackage{subfigure}
\usepackage{amsmath,amssymb,amsfonts}
\title{Research on Resource Allocation under Unlicensed Spectrum Using Q-Learning
}

\author{
  Uyoy Ial \\
   \\
  Department of Engineering Science, UKCN \\
  \texttt{16046043b@gmail.com} \\
}

\begin{document}
\maketitle

\begin{abstract}
In response to the advent of the 5G era, enhancing throughput and increasing transmission efficiency within limited spectrum resources is an important research topic. In the LTE system, utilizing unlicensed spectrum to assist traditional mobile networks, known as License Assisted Access, has emerged as a viable solution to effectively improve transmission efficiency. However, as the unlicensed spectrum also accommodates other users, such as Wi-Fi for mobile communication, there is a need to address the issue of spectrum resource allocation, aiming to achieve fair transmission among different mobile communication users. This research project aims to explore and compare two approaches: traditional communication algorithms and reinforcement learning method Q-leaning, under the condition of achieving maximum system throughput, in order to determine the differences between the two methods.

\end{abstract}


\section{Introduction}
Spectrum resources stand as both incredibly scarce and immensely valuable assets in the domain of telecommunications. With the ever-increasing demand for wireless connectivity and the surging volume of data traversing mobile networks, the available bandwidth in licensed spectrum is progressively becoming inadequate to meet the burgeoning requirements. This scenario presents a pressing challenge for telecommunication providers and necessitates innovative solutions to address the impending spectrum scarcity issue.

Enter the utilization of unlicensed spectrum—an avenue that offers a promising solution to mitigate the strains on licensed spectrum resources. As the demands placed on networks continue to soar, meeting the critical benchmarks set forth by 5G technology becomes non-negotiable. The quintessential goals of 5G, including significantly swifter data transmission, bolstered reliability, and substantially reduced latency, have emerged as pivotal requirements to support the growing ecosystem of connected devices, IoT applications, and bandwidth-hungry services.

However, in light of the limitations imposed by finite spectrum resources, expanding the capabilities of wireless networks becomes imperative. This challenge has spurred the development of innovative technologies such as License Assisted Access (LAA) \cite{ratasuk2014lte}, which leverages Carrier Aggregation (CA) \cite{yuan2010carrier} to harness the potential of unlicensed bands alongside licensed ones. LAA acts as a catalyst, empowering data transmission in unlicensed bands to complement and fortify the data throughput and speed available in the licensed spectrum.

One of the groundbreaking aspects of LAA is its ability to facilitate LTE users' access to the unlicensed 5 GHz band \cite{yin2016framework}, effectively allowing seamless sharing of its resources with Wi-Fi networks. This harmonious sharing not only enhances the overall efficiency of spectrum usage but also optimizes the spectrum's potential, ensuring a more robust and inclusive network environment.

The successful implementation of LAA in the realm of 4G technology has laid a sturdy groundwork, providing a blueprint for effectively utilizing unlicensed spectrum resources. This success story serves as a guiding beacon for standardization bodies like 3GPP \cite{3gpp2015study} and telecommunications operators, offering invaluable insights for the formulation and promotion of cutting-edge standards such as 5G NR-U. These standards will shape the future landscape of telecommunications, fostering more efficient spectrum utilization and enabling the realization of enhanced mobile connectivity in the 5G era and beyond.

In this paper, our objective is to devise a reinforcement learning approach, specifically a Q-learning-based channel selection algorithm, alongside a traditional minimum interference distance-based channel selection method. Our goal is to enhance the efficiency of unlicensed spectrum resource allocation by integrating a spectrum map. We aim to compare and analyze the disparities in throughput and characteristics brought about by these algorithms.



\section{Reinforcement-Learning}
Reinforcement Learning (RL) \cite{kaelbling1996reinforcement} represents a pivotal domain within the realm of machine learning, focusing on the ability of an agent to make sequential decisions through interactions with an environment. This learning paradigm is rooted in the concept of learning from experience, where an agent learns by taking actions, receiving feedback, and adapting its strategies over time to maximize cumulative rewards.

\begin{itemize}

\item Policy: The policy in RL embodies the strategy or set of rules that the agent employs to select actions based on the observed state of the environment. It essentially maps states to actions, aiming to optimize long-term rewards.

\item Reward: Rewards serve as the feedback mechanism provided by the environment in response to the actions taken by the agent. They indicate the immediate benefit or cost associated with a particular action in a specific state. The goal of the agent is to learn a policy that maximizes the cumulative reward over time.

\item Value Function: The value function estimates the long-term value or utility of being in a particular state and following a specific policy. It represents the expected cumulative reward an agent anticipates by starting from a given state and following a particular strategy.

\item Environment Model: In RL, the environment model is the framework within which the agent operates. The agent interacts with the environment by taking actions, receiving rewards, and transitioning to new states based on these actions. The model doesn't always need to be explicitly known to the agent; it can learn directly from experience through interaction.

\end{itemize}

RL systems strive to learn optimal policies by exploring different actions in various states of the environment, taking into account the received rewards and using them to update value functions or policies iteratively. RL has found extensive applications in telecommunications \cite{maglogiannis2018q}, \cite{rastegardoost2018machine}, and \cite{tan2020intelligent}. Furthermore, its utilization in next-generation wireless communication resource allocation issues, such as 5G, has shown promise. RL methods are adept at addressing wireless communication scenarios where the environment model is unknown or where uncertainty exists in understanding the clear state of the environment.

\section{Q-Learning}
\begin{figure}[h]
	\centerline{\includegraphics[width = 0.35\linewidth]{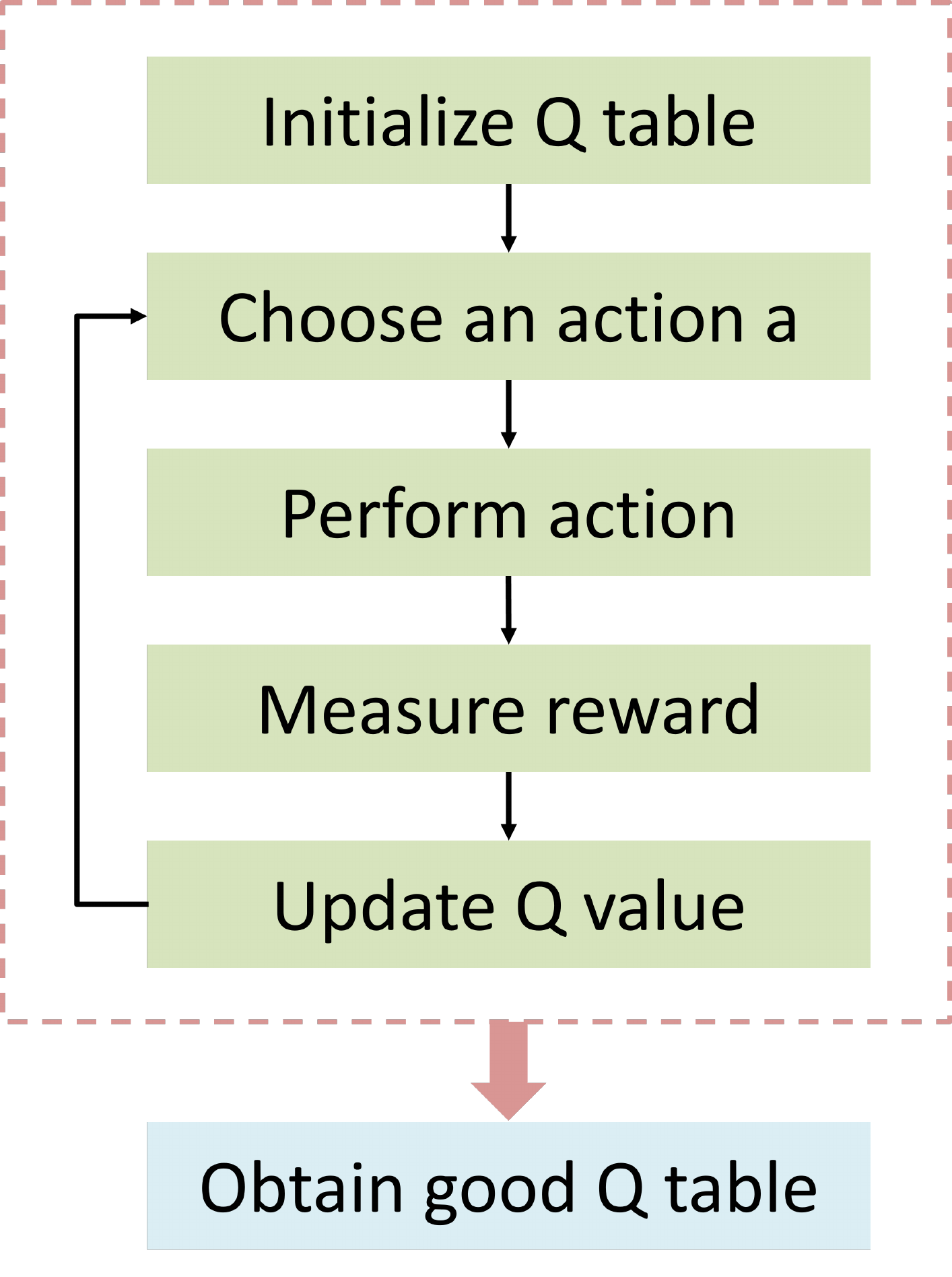}}
	\caption{Q-Learning algorithm flow chart.}
	\label{fig:q-learning}
\end{figure}

Q-Learning \cite{watkins1992q} is one of the methods in Reinforcement Learning, where an unsupervised agent perceives the environment, generates corresponding feedback, and learns to choose actions that lead to optimal rewards. In other words, without the need for prior knowledge of the environment model, Q-Learning selects strategies by learning action value functions within any given Finite Markov Decision Process. The algorithmic flow of Q-Learning is shown in Figure \ref{fig:q-learning}.

The updating principle for the Q-table is mathematically represented as:
\begin{equation}
    Q(s, a) \leftarrow (1 - \alpha) \cdot Q(s, a) + \alpha \cdot (r + \gamma \cdot \max Q(s', a')),
\end{equation}
where $Q(s, a)$ represents the old Q value function, $\alpha$ denotes the learning rate, typically between 0 and 1, determining the importance of new information in updating the old values, $r$  signifies the reward obtained, $\gamma$ is the discount factor, within the range [0, 1], indicating the significance of future rewards, and $s'$ and $a'$ denote the next state and action respectively. $\max Q(s', a')$ estimates the optimal future Q value from the next state.

DQN (Deep Q-Network) \cite{mnih2015human}, Double DQN \cite{van2016deep}, and Dueling DQN \cite{wang2016dueling} are advanced variations and extensions of the Q-Learning algorithm, particularly designed to address the challenges of learning in complex environments with high-dimensional state spaces, such as those encountered in reinforcement learning tasks.

\subsection{DQN (Deep Q-Network)}
DQN introduces the use of deep neural networks as function approximators to estimate Q-values, enabling learning in high-dimensional state spaces.
It employs an experience replay mechanism, storing past experiences (transitions) in a replay buffer and randomly sampling mini-batches from this buffer to break the correlations between consecutive experiences. This helps stabilize learning and improves data efficiency.
DQN also utilizes a separate target network, which is a copy of the primary network but with delayed updates. This stabilizes training by fixing the target Q-values for several iterations before updating them.

The basic Q-learning formula for DQN is:
\begin{equation}
    Q(s, a) \leftarrow (1 - \alpha) \cdot Q(s, a) + \alpha \cdot \left(r + \gamma \cdot \max Q(s', a')\right).
\end{equation}

\subsection{Double DQN (Double Deep Q-Network)}
Double DQN  addresses the overestimation bias inherent in traditional Q-learning methods, particularly in DQN, where the maximum Q-value for the next state is used for action selection and evaluation.
It introduces the concept of using two separate neural networks: one for action selection (policy network) and another for Q-value estimation (value network). This mitigates the potential overestimation of Q-values by decoupling the action selection from the value estimation.

The formula for Double DQN, based on DQN, involves using a separate neural network to evaluate the Q-value of the best action in the next state $s'$, reducing the overestimation of Q-values:

\begin{equation}
    Q(s, a) \leftarrow (1 - \alpha) \cdot Q(s, a) + \alpha \cdot \left(r + \gamma \cdot Q_{\text{target}}\left(s', \arg\max Q(s', a')\right)\right),
\end{equation}
where $Q_{\text{target}}$ is another network used to calculate the target Q-value.

\subsection{Dueling DQN (Dueling Deep Q-Network)}
Dueling DQN separates the Q-value function into two streams: one estimating the value of being in a particular state (state value) and another estimating the advantage of each action over the others.
By decoupling the estimation of state values and action advantages, Dueling DQN aims to improve learning efficiency by learning which states are valuable and which actions are advantageous independently.
This architecture enables the agent to focus more on relevant state features and actions, leading to more effective learning in reinforcement learning tasks.

Dueling DQN aims to separate the Q-value function into state value and action advantages. Its formula is:
\begin{equation}
    Q(s, a) \leftarrow V(s) + A(s, a) - \frac{1}{|\mathcal{A}|} \sum_{a'} A(s, a'),
\end{equation}
where $V(s)$ represents the state value, $A(s, a)$ is the action advantage for each action $a$., and $|\mathcal{A}|$
 is the number of possible actions.
 
\section{Research Framework}
\begin{figure}[h]
	\centerline{\includegraphics[width = 0.5\linewidth]{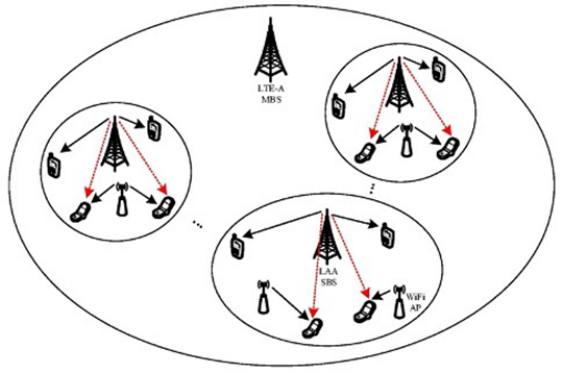}}
	\caption{System architecture diagram.}
	\label{fig:framework_LAA}
\end{figure}

Our research framework, including the environment setup and system architecture. The system is centered around a macro base station (MBS), which randomly generates small base stations (SBS) within its range. Subsequently, within the coverage area of the SBS, there are generated entities such as LAA users (LAA UE), WiFi access points (WiFi AP), and WiFi users (WiFi UE). The system architecture diagram is shown in Figure \ref{fig:framework_LAA}

Within this system, LAA UE experiences interference from the environment, leading to reduced transmission efficiency. The interference is primarily categorized into two types: interference caused by WiFi AP on LAA UE ($I_{\text{AP\_to\_UE}}$) and interference among LAA UE themselves ($I_{\text{UE\_to\_UE}}$). The premise for interference is the use of the same frequency channel. The goal of both the minimum interference distance-based channel selection algorithm and Q-learning is to select the optimal channel for users, aiming to maximize overall throughput.

Throughput is calculated using the following formula:

$
    \text{Num} = P_{\text{LAA}} \times \text{fading\_gain\_MBS\_LAAUE},\\
    \text{Deno} = I_{\text{AP\_to\_UE}} + I_{\text{UE\_to\_UE}} + N_o,\\
$

\begin{equation}
    \text{Throughput} = \frac{T_{\text{max}} \times \log_{10}(1 + \frac{\text{Num}}{\text{Deno}})}{I_{\text{cca}} + T_{\text{max}}},\\
\end{equation}

where $P_{\text{LAA}} = P_{\text{MAX}} = 10^{-3} \times 10^{\frac{24}{10}}$, $T_{\text{max}} = 10 \text{ ms}$, $
I_{\text{cca}} = 0.0034$, $\text{ ms}$, $
N_o = 2 \times 10^{-13}$.

The values for $\text{fading\_gain\_MBS\_SBSUE}$, $I_{\text{AP\_to\_UE}}$, and $I_{\text{UE\_to\_UE}}$ are calculated using additional formulas \cite{9013885}.

\section{Method}
We utilized two distinct methods: the traditional Minimum Interference Distance (MID) Channel Selection Method and the reinforcement learning-based Q-learning Channel Selection Method:
\begin{itemize}

    \item MID Channel Selection Method
    
    This method prioritizes the selection of idle channels for transmission. In cases where there are no available idle channels, it calculates the distance between users' coordinates. By selecting channels with greater distances that result in less interference, it aims to enhance transmission efficiency.

    \item Q-learning Channel Selection Method
    
    This method involves inputting the coordinates of users within the environment along with the channels they are using. It employs Q-learning techniques, using Throughput as its reward, aiming to maximize the Throughput value. The information of users within the environment is fed into the trained final model, allowing the model to select the optimal channel for the current scenario.
    
\end{itemize}

\section{Experiment Results}
\begin{figure}[h]
	\centerline{\includegraphics[width = 0.5\linewidth]{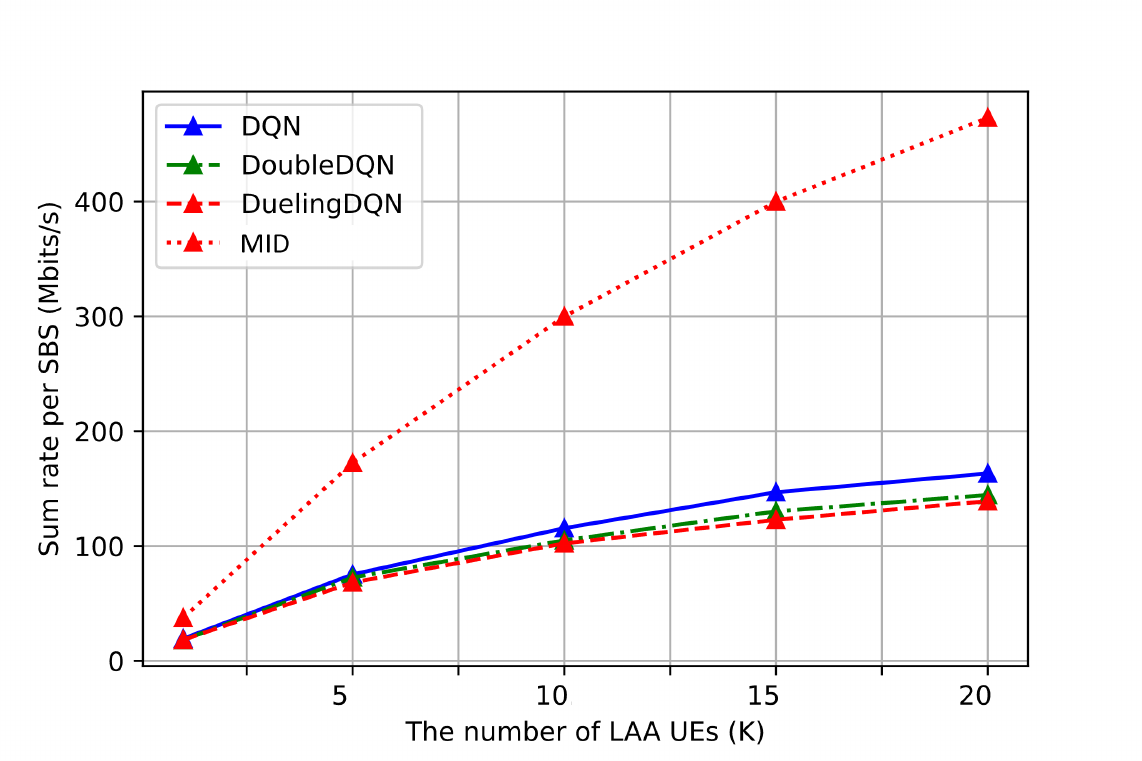}}
	\caption{Throughput analysis chart for varying numbers of LAA UEs.}
	\label{fig:laa_ue}
\end{figure}

\begin{figure}[h]
	\centerline{\includegraphics[width = 0.5\linewidth]{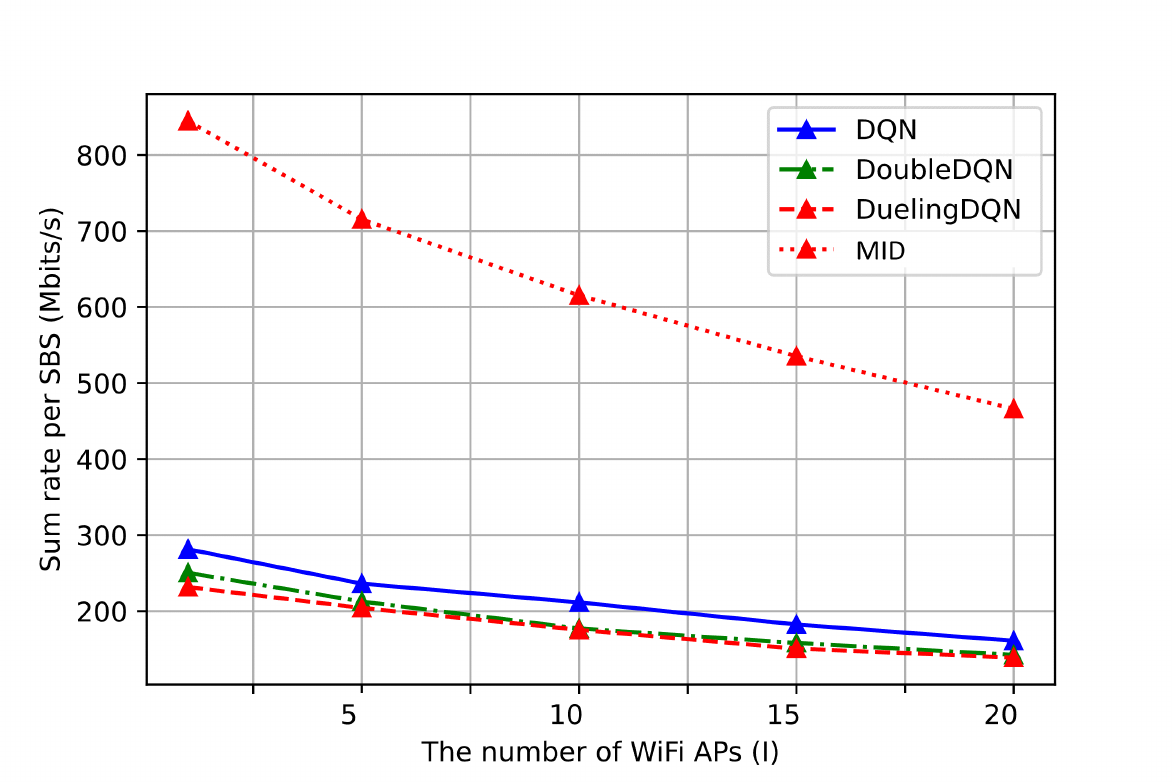}}
	\caption{Throughput analysis chart for varying numbers of WIFI APs.}
	\label{fig:wifi_ap}
\end{figure}

In our experiments, we employed a 5-layer fully connected neural network as the Q-learning network. The number of neurons in each layer decreases with depth. We set the value of $\alpha$ to 0.05 and the value of $\gamma$ to 0.9. In the experiment, we set the number of channels to 15, and we conducted experiments with fixed quantities of WiFi AP and LAA UE separately. We compared the traditional MID Channel Selection Method with three Q-learning Selection Methods, and the results are displayed in Figure \ref{fig:laa_ue} and Figure \ref{fig:wifi_ap}. 

During the experiments, after 20,000 training iterations, it was observed that there still existed a considerable gap in performance between the three Q-learning methods and the MID Channel Selection Method. This disparity might be attributed to the inadequate parameters of the neural network utilized in the Q-learning models, limiting the model's capacity. Additionally, the performance differences among DQN, Double DQN, and Dueling DQN were minimal, potentially due to constraints imposed by the model architecture. Although Double DQN and Dueling DQN were designed to alleviate the overestimation of Q-values observed in DQN, this phenomenon was not prevalent in our experiments.

In future endeavors, the aim is to identify more suitable Q-learning model architectures, such as CNN or Transformer models. Furthermore, an interesting observation was made regarding the MID Channel Selection Method's computational complexity, which scales as $O(n^2)$ as the number of users increases. This suggests that in more complex scenarios, Q-learning might have a better chance to showcase its advantages. Nevertheless, achieving the expected level of performance in Q-learning requires a well-suited model structure and appropriate training.

\section{Conclusion}
In this paper, we conducted a comparison between the traditional MID Channel Selection Method and various Q-learning selection algorithms for wireless resource allocation in communication systems. Through experimentation and analysis, a significant performance gap between Q-learning and the established MID method persists. Potential reasons for this disparity include limitations imposed by the neural network parameters in the Q-learning models, possibly constraining the overall learning capability. Furthermore, observed differences in performance among the tested Q-learning variants (DQN, Double DQN, and Dueling DQN) were minimal. We also noted that in more complex scenarios, Q-learning may have greater potential to showcase its advantages. Our future work will focus on exploring more suitable Q-learning model architectures, such as CNN or Transformer models, to enhance the learning capacity in wireless resource allocation scenarios.

\bibliographystyle{unsrt}  
\bibliography{references}

\begin{thebibliography}{10}

\bibitem{ratasuk2014lte}
Rapeepat Ratasuk, Nitin Mangalvedhe, and Amitava Ghosh.
\newblock Lte in unlicensed spectrum using licensed-assisted access.
\newblock In {\em 2014 IEEE Globecom Workshops (GC Wkshps)}, pages 746--751. IEEE, 2014.

\bibitem{yuan2010carrier}
Guangxiang Yuan, Xiang Zhang, Wenbo Wang, and Yang Yang.
\newblock Carrier aggregation for lte-advanced mobile communication systems.
\newblock {\em IEEE communications Magazine}, 48(2):88--93, 2010.

\bibitem{yin2016framework}
Rui Yin, Guanding Yu, Amine Maaref, and Geoffrey~Ye Li.
\newblock A framework for co-channel interference and collision probability tradeoff in lte licensed-assisted access networks.
\newblock {\em IEEE Transactions on Wireless Communications}, 15(9):6078--6090, 2016.

\bibitem{3gpp2015study}
3GPP.
\newblock Study on licensed-assisted access to unlicensed spectrum.
\newblock {\em 3rd Generation Partnership Project (3GPP)}, 6:TR--36, 2015.

\bibitem{kaelbling1996reinforcement}
Leslie~Pack Kaelbling, Michael~L Littman, and Andrew~W Moore.
\newblock Reinforcement learning: A survey.
\newblock {\em Journal of artificial intelligence research}, 4:237--285, 1996.

\bibitem{maglogiannis2018q}
Vasilis Maglogiannis, Dries Naudts, Adnan Shahid, and Ingrid Moerman.
\newblock A q-learning scheme for fair coexistence between lte and wi-fi in unlicensed spectrum.
\newblock {\em IEEE Access}, 6:27278--27293, 2018.

\bibitem{rastegardoost2018machine}
Nazanin Rastegardoost and Bijan Jabbari.
\newblock A machine learning algorithm for unlicensed lte and wifi spectrum sharing.
\newblock In {\em 2018 IEEE International Symposium on Dynamic Spectrum Access Networks (DySPAN)}, pages 1--6. IEEE, 2018.

\bibitem{tan2020intelligent}
Junjie Tan, Lin Zhang, Ying-Chang Liang, and Dusit Niyato.
\newblock Intelligent sharing for lte and wifi systems in unlicensed bands: A deep reinforcement learning approach.
\newblock {\em IEEE Transactions on Communications}, 68(5):2793--2808, 2020.

\bibitem{watkins1992q}
Christopher~JCH Watkins and Peter Dayan.
\newblock Q-learning.
\newblock {\em Machine learning}, 8:279--292, 1992.

\bibitem{mnih2015human}
Volodymyr Mnih, Koray Kavukcuoglu, David Silver, Andrei~A Rusu, Joel Veness, Marc~G Bellemare, Alex Graves, Martin Riedmiller, Andreas~K Fidjeland, Georg Ostrovski, et~al.
\newblock Human-level control through deep reinforcement learning.
\newblock {\em nature}, 518(7540):529--533, 2015.

\bibitem{van2016deep}
Hado Van~Hasselt, Arthur Guez, and David Silver.
\newblock Deep reinforcement learning with double q-learning.
\newblock In {\em Proceedings of the AAAI conference on artificial intelligence}, volume~30, 2016.

\bibitem{wang2016dueling}
Ziyu Wang, Tom Schaul, Matteo Hessel, Hado Hasselt, Marc Lanctot, and Nando Freitas.
\newblock Dueling network architectures for deep reinforcement learning.
\newblock In {\em International conference on machine learning}, pages 1995--2003. PMLR, 2016.

\bibitem{9013885}
Yi-Feng Huang and Hsiao-Hwa Chen.
\newblock On sum-rate maximization in cr-assisted heterogeneous lte-laa networks.
\newblock In {\em 2019 IEEE Global Communications Conference (GLOBECOM)}, pages 1--6, 2019.

\end{thebibliography}

\end{document}